\documentclass{stanfest}
\author{Richard Ignace}[ETSU]
\affil[ETSU]{Department of Physics \& Astronomy,
East Tennessee State University, Johnson City, Tennessee, USA}

\title{Interpretative Modeling of Structured Winds 
Using Linear Polarization}

\begin{document}

\maketitle

\begin{abstract}

Polarization provides additional diagnostic opportunities for probing
the structured environments of massive stars as well as the
illumination of those environments by stars that are not spherical.
After a brief overview of polarization considerations relevant to
hot massive stars, selected applications are presented.  Examples
related to dense Wolf-Rayet winds are chosen:  clumpy wind flow,
co-rotating interaction regions, and colliding wind interactions.
Brief remarks are given about the prospects for opening a new window
on massive star studies using UV spectropolarimetry with the {\em
Polstar} mission concept.

\end{abstract}

\section{Introduction}
\label{sec:intro}

Astrophysical polarization is a broad topic with varied applications
that can be source and waveband specific.  Polarization can arise
from scattering, the Zeeman effect, synchrotron emission; interstellar
polarization is associated with extinction effects involving grain
alignment.  Linear polarimetry of unresolved stars has proven to
be an important tool for diagnosing source geometery
\citep[e.g.,][]{2010stpo.book.....C}.  This is because spherical
symmetry yields no net polarization, so that detection of intrinsic
polarization signifies a deviation from spherical.  Such symmetry
breaking can come from stellar distortion (e.g., rotation),
non-spherical circumstellar media (e.g., Be stars), or multiplicity
of stars (e.g., eclipses).  The reader is referred to the monographs
by \cite{2004ASSL..307.....L} and \cite{2010stpo.book.....C} for a
comprehensive review of the history, methods, and applications of
polarimetry to stars.  Three specific applications are considered
in relation to the structured winds of massive stars.

\section{Modeling Linear Polarization from Scattering by Electrons}
\label{sec:modeling}

The focus of this contribution is on hot massive stars with highly
ionized winds for which the dominant continuum polarigenic opacity
will be dipole (Thoomson) scattering by
free electrons.
The description of polarized radiation typically follows
the notation of the Stokes parameters $I, Q, U,$ and $V$.  Note these 
are frequency (wavelength) dependent 
intensities.  Stokes-I refers to the total intensity; Q and U are
measures of linear polarization;
and Stokes-V is a measure of the circular
polarization.  In what follows only linear polarization with
described, so $V=0$.

\cite{1960ratr.book.....C} summarizes the radiative transfer for
dipole scattering.  He also uses 
an alternative notation of $I_l$, $I_{\rm r}$, U, and V.
Here the ``l'' and ``r'' refer to intensities associated with the
plane of scattering, either in the plane, or orthogonal to it, and
$I=I_l+I_{\rm r}$ and $Q=I_l-I_{\rm r}$.  For pure scattering the
source functions are

\begin{equation}
S_l =  \frac{3}{4}\,\left[ 2(J_l-K_l) + \cos^2\chi \, (3K_l -2J_l +J_r)\right] ;\;
S_r =  \frac{3}{4}\,\left(J_r+K_l\right) 
\end{equation}

\noindent where $J$ and $K$ refer to appropriate Eddington moments
\citep[e.g.,][]{2014tsa..book.....H}, and $\chi$ of the angle of scattering.

What complicates the scattering problem is that the J and K quantities
depend on the source functions for all other points, which is the
standard challenge of solving an integro-differential equation for
the emergent intensities.  However, simplifications
result when assuming (a) the scattering circumstellar medium is
optically thin, (b) the star can be treated as a point source, and
(c) the incident radiation at the location of scattering is completely
unpolarized.  These lead to $J_l = J_{\rm r} = J/2$ and $
K_l=K_{\rm r}=K/2$, for $J$ and $K$ the Eddington moments
of a uniformly bright star.  For a point source, $J=K=W(r)I_\ast$,
for $I_\ast$ the isotropic intensity at the stellar surface. Additionally,
the dilution factor becomes $W(r) \rightarrow
\frac{1}{4}\,\frac{R_\ast^2}{r^2}$.  Then

\begin{eqnarray}
S_I & = & I_{\rm r}+I_l = \frac{3}{8}\,\left[(3J-K) + (3K-J)\cos^2\chi\right]
	\approx \frac{3}{16}\,I_\ast\,\frac{R^2}{r^2}\,(1+\cos^2\chi), \\
S_Q & = & I_{\rm r} - I_l = \frac{3}{8}\,(3K-J)\,(1-\cos^2\chi) \approx
	\frac{3}{16}\,I_\ast\,\frac{R^2}{r^2}\,\sin^2\chi.
\end{eqnarray}

\noindent For the scattering plane oriented about the observer axis by angle
$\psi$, the emergent Stokes intensities will be

\begin{equation}
I = \int\,n_{\rm e}\,\sigma_T\,S_I\,dz ;\;
Q = \int\,n_{\rm e}\,\sigma_T\,S_Q\,\cos 2\psi\,dz ;\;
U = \int\,n_{\rm e}\,\sigma_T\,S_Q\,\sin 2\psi\,dz ;
\end{equation}

\noindent where $n_{\rm e}$ is the number density of electrons,
$\sigma_T$ is the Thomson cross-section, and $z$ is the observer
sightline at cylindrical impact parameter $\varpi$, and angle $\psi$.
This is a standard convention indicating a sequence of signs for
polarization such that $(Q,U)$ are $(+,0), (0,+), (-,0), (0,-)$ for
azimuths of $\psi = 0^\circ, 45^\circ, 90^\circ, 135^\circ$.  

The stellar monochromatic specific intensity for a uniformly bright
star is implicitly $L_\ast = 4\pi^2\,R^2\,I_\ast$.  Combining with
the point source expressions for $S_I$ and $S_Q$, and integrating
over all rays, the following Stokes luminosities for scattered light
are obtained:

\begin{eqnarray}
L_I & = & \frac{3}{16\pi}\,L_\ast\,\int\,n_{\rm e}\,\sigma_T\,(1+\cos^2\chi)\,\frac{dV}{r^2} \\
L_Q & = & \frac{3}{16\pi}\,L_\ast\,\int\,n_{\rm e}\,\sigma_T\,\sin^2\chi\,\cos 2\psi\,\frac{dV}{r^2} \\
L_U & = & \frac{3}{16\pi}\,L_\ast\,\int\,n_{\rm e}\,\sigma_T\,\sin^2\chi\,\sin 2\psi\,\frac{dV}{r^2} ,
\end{eqnarray}

\noindent where $dV = \varpi \, d\varpi\, d\psi\,dz$ as expressed
in observer coordinates.  However, the volume element may be expressed
in whatever coordinates are most convenient, such as defined in the
stellar system.  

These expressions for scattered light properties involve volume
integrals owing to the source being unresolved.  But the expressions
are moment equations involving scattering angle $\chi$ and orientation
$\psi$ as weighted by the electron density, and thus encode within
them structure information about the scattering medium despite being
spatially unresolved \citep[e.g.,][]{1982MNRAS.200...91S,
1983MNRAS.205..153S}.  Not considered here, similar considerations
arise when the illuminating source is not isotropic
\citep[e.g.,][]{1999A&A...347..919A, 2009A&A...496..503I}. There
can also be a distribution of illuminating sources, such as multiple
stars \citep[e.g.,][]{1978A&A....68..415B} or diffuse emission from
the medium itself.

The preceding 3 expressions are typically normalized to form relative
quantities.  Assuming $L_I \ll L_\ast$, then
relative quantities in the relatively traditional form result:

\begin{equation}
f_{\rm sc} = L_I/L_\ast ;\;
q = L_Q/L_\ast ;\;
u = L_U/L_\ast ;\;
p = \sqrt{q^2+u^2} ;\;
\tan 2 \psi_{\rm p} = u/q;
\end{equation}

\noindent where subscript ``sc'' makes explicit the scattered light
nature for the Stoke-I fluctuations, lower case $q$ and $u$ are
fractional Stokes polarizations, $p$ is the total polarization and
positive definite, and the $180^\circ$ ambiguity in the PA, $\psi_{\rm
p}$, is apparent through the form of its definition.  Key is that
sign changes in either of $q$ or $u$ amount to PA ``flips'' of
$90^\circ$ which is a valuable diagnostic for geometry in
unresolved sources, relating either to evolving geometry or
to wavelength-dependent opacity influences.

There are two more versions
of these expressions worth considering.  The first involves
re-expressing the density in terms of a structure function, using
$n_{\rm e} = n_0\,G(t,r,\theta,\phi)$ for $n_0$ some reference
density in the envelope, $(r,\theta,\phi)$ or $(r,\mu,\phi)$ with
$\mu=\cos\theta$ as convenient stellar coordinates, and $G$ a
``structure'' equation for geometry of the scattering density.  In
effect, $G$ encodes deviations from sphericity that can be probed
by the polarization observables, which could include time variability.
Using the normalized forms, the scattered light properties become

\begin{eqnarray}
f_{\rm sc} & = & \frac{3}{16\pi}\,\tau_0\,\int\,G(t,\tilde{r},\mu,\phi)\,(1+\cos^2\chi)\,d\tilde{r}\,d\mu\,d\phi \\
q & = & \frac{3}{16\pi}\,\tau_0\,\int\,G(t,\tilde{r},\mu,\phi)\,\sin^2\chi\,\cos 2\psi\,d\tilde{r}\,d\mu\,d\phi \\
u & = & \frac{3}{16\pi}\,\tau_0\,\int\,G(t,\tilde{r},\mu,\phi)\,\sin^2\chi\,\sin 2\psi\, d\tilde{r}\,d\mu\,d\phi,
\end{eqnarray}

\noindent where a normalized radius is introduced, $\tilde{r} =
r/R$, and $\tau_0 = n_0\sigma_T R$.  Consider these expressions as
the product of two parts.  First are the coefficients outside the
respective integrals.  The produce $n_0\sigma_T R$ represents an
optical depth scale of the envelope.  The integrals represent spatial
averaging the structure function $G$ as weighted by the mode of
polarigenic opacity.  In this case the latter is dipole scattering
as signified by factors involving the scattering angle $\chi$.  If
the scattering were Rayleigh, the expressions are the same except
for the introduction of wavelength dependence.  If dust, and optically
thin, the scattering portion and chromatic infuence would be
different, but the weighting by $G$ would remain.

Second is the axisymmetric scenario.  In this case $G=G(t,r,\mu)$.
We take the symmetry axis of the star to be inclined to the observer
axis by angle $i$.  Spherical trig relations map observer angles
$(\chi,\psi)$ to stellar angles $(\theta, \phi)$.  One of these
is $\cos \chi = \cos \theta \cos i+\sin\theta\sin i\cos(\phi)$.
Laws of Sines combined with Law of Cosines for the $\theta$ arc
allows to perform the integration in $\phi$.  The classic result
of \cite{1977A&A....57..141B} follows with

\begin{eqnarray}
f_{\rm sc} & = & \frac{3}{16}\,\tau_0\,\int\,G(t,\tilde{r},\mu)\,\left[(2+\sin^2 i) +
	(2-3\sin^2 i)\,\mu^2\right]\,d\tilde{r}\,d\mu \\ 
q & = & \frac{3}{16}\,\tau_0\,\sin^2 i\,\int\,G(t,\tilde{r},\mu)\,(1-3\mu^2)\,
	d\tilde{r}\,d\mu\\
u & = & \frac{3}{16}\,\tau_0\,\sin^2 i\,\int\,G(t,\tilde{r},\mu)\,(1-3\mu^2)\,
	d\tilde{r}\,d\mu = 0,
\end{eqnarray}

\noindent There are several points of interest.  (1) By symmetry,
$u=0$.  This assumes the projected stellar axis is aligned with the
measurement system. If not the case, the observer measurements
$(q',u')$ will be $(q\cos 2\psi_{\rm p}, q\sin 2\psi_{\rm p})$,
where $\psi_{\rm p}$, where $\psi_{\rm p}$ is the azimuth between
the observer system and the stellar symmetry axis on the sky.  (2)
The polarization scales with $\sin^2 i$ and vanishes if the star
is observed pole-on.  When pole-on, the scattered light is
centro-symmetric on the sky, and there is exact cancellation in $q$
and $u$.  (3) There is scattered light for all inclinations.  (4)
Finally, the function $G$ need not be top-bottom symmetric.

The axisymmetric solutions have tremendous utility, when the various
assumptions have been met.  For example, the inclination $i$ could
be time dependent.  For example, \cite{2013ApJ...766L...9C} modeled
the variable polarization for magnetospherically confined plasma
in the Bp star $\sigma$~Ori~E as with a ``dumbbell'' distribution.
In this model $i=i(t)$ as a function of rotation phase.  Additionaly
for thin scattering, one can configure a distribution of axisymmetric
structures through superposition.  Now each structure will have a
set of $\{ i_{\rm k}\}$ and $\{ \psi_{\rm k}\}$.
\cite{1996A&A...306..519R} and \cite{2000A&A...357..233L} did this
for clumps, and \cite{2015A&A...575A.129I} configured a CIR this
way.

\cite{1978A&A....68..415B} have also expanded the analysis to include
an arbitrary number of point sources, with a general analysis
relevant to binary stars.  Improvements to the illumination have
been considered by \cite{1987ApJ...317..290C} and
\cite{1989ApJ...344..341B} to relax the point source assumption and
allow for limb darkening of the star.  One returns to the expressions
for Eddington moments.  \cite{1994ApJ...436..818B} developed an
approach for including gravity darkening using generalized Eddington
moments.

\section{Applications}
\label{sec:apps}

Each of the example applications involve Wolf-Rayet (WR) stars.
Classical WR stars \citep[cWRs;][]{2007ARA&A..45..177C} come in 4
main types:  N-enriched but with some H, WNh; N-rich and H absent,
WN; C-rich, WC; and O-rich, WO.  The cWRs are understood to be
endstates of evolution for massive stars of sufficient mass
\citep[e.g.,][]{2012ARA&A..50..107L}.  

\subsection{Clumpy Winds}

There is ample evidence to support that cWR winds are clumpy
\citep[e.g., see][]{2008A&ARv..16..209P}.
Among the studies are polarimetric ones revealing $q-u$
scatter plots as expected from a stochastic process \citep{1987ApJ...322..870S,
1987ApJ...322..888D}.  Here ``scatter plot'' means that time averages
of $\langle q \rangle \approx \langle u\rangle \approx 0$.  As a result,
$\psi_{\rm p}$ is random meaning no preferred
axis for the polarization.  Such results are consistent with clumpy
wind flows that are spherical in time average.  Yet the average
polarization $\langle p \rangle > 0$,
and both the photometric and polarimetric light curves yield measures
of standard deviation $\sigma_{\rm m}$ (here, ``m'' for magnitudes)
and $\sigma_{\rm p}$.  These 3
observables can be used to constrain the properties of the clumps,
such as their masses and injection rate.

Several have attempted to model the observations.  In particular,
\cite{1995A&A...295..725B} and \cite{1996A&A...306..519R} developed
models for large numbers of randomly distributed (small) thin
scattering clumps using expressions from the preceding section.
Their approach does not reproduce the observed ratio of
$\sigma_{\rm}/\sigma_{\rm p} \sim 20$, being about $3\times$ too
small.  They assumed strictly thin scattering whereas WR winds are
optically thick, and those authors point out that inclusion of multiple
scattering and/or free-free/bound-free emission could help to
reproduce the observations.

Some have considered multiple scattering in clumps
\citep{1995ApJ...441..400C, 2000ApJ...540..412R, 2007A&A...469.1045D,
2012AIPC.1429..278T}.  However, these have attempted to evolve
clumps as a flow to simulate light curves.  Taking a different
approach, \cite{2019MNRAS.490.5921R} adopted a simple time function
to model the variable light curve of the WN8 star WR~40 as observed
by BRITE.  The time function is a half-gaussian of the form,

\begin{equation}
L(t) = \sum\, L_{\rm j} \, e^{-(t-t_{\rm j})^2/\tau^2},
\end{equation}

\noindent where $t_{\rm j}$ is the time of injection of a clump in
the flow, $\tau$ is timescale for the decline in clump brightness,
and $L_{\rm j}$ is a luminosity amplitude.  The summation is over
all clumps for which $t_{\rm j}<t$. This functional form reasonably
assumes that a clump is brightest when appearing at the WR's
pseudo-photosphere and dims monotonically with time, as it associated
with moving out in the flow.

\cite{2019MNRAS.490.5921R} could tune model parameters to replicate
the observed photometric fluctuations in the 130~d lightcurve.  An
independent dataset for WR~40 involving polarization was then tested
by \cite{2023MNRAS.519.3271I}.  Although more limited, the photometry
and polarimetry were simultaneous, and inclusion of polarimetry
provided new constraints in the form of $\langle p\rangle$ and
$\sigma_{\rm p}$.  A new time function was introduced to represent
the polarization, such that $p=0$ at the moment of injection \citep[as
expected from][]{1987ApJ...317..290C}. Although again reproducing
the characteristics of the data, distinguishing between equal clumps
versus a clump distribution was not possible.

\subsection{Co-Rotating Interaction Regions}

The era of space-based ultraviolet (UV) spectroscopy led to the
discovery of strong P~Cygni lines for massive star, indicative of
expanding winds \citep{1976ApJS...32..715L}.  Observations for a
large number of O~stars revealed discrete absorption compnents
\citep[``DACs,][]{1999A&A...344..231K, 2015ApJ...809...12M}.
\cite{1984ApJ...283..303M} suggested these spectral features might
be explained in terms of co-rotating interaction regions (CIRs).
\cite{1996ApJ...462..469C} explored this with 2D hydrodynamics to
find good matches to the observed features of DACs.  Subsequent
modeling and applications have supported the CIR hypothesis
\citep{2004A&A...423..693D, 2008ApJ...678..408L, 2017MNRAS.470.3672D}.

The presence of CIRs in WR winds has also been inferred
\citep{2009ApJ...698.1951S, 2011ApJ...735...34C, 2011ApJ...736..140C,
2016MNRAS.460.3407A}.  Recent linear polarimetric data reveal
unexpected features across optical emission lines in the WN4 star
WR~6 (EZ~CMa), as seen in Fig.~1 of
\cite{2023MNRAS.526.1298I}.  Moving from the highest redshift velocities
toward the line center, polarization is seen to decrease.  This is
the classic ``line effect'', whereby line photons if not scattered
contribute to the normalization of the polarization and act
to ``dilute'' the intrinsic continuum polarization
across the line. 

However, from line center through blueshifted velocities, the
polarization forms a loop in the $q-u$ diagram.  One way to explain
this behavior is with a CIR model.  When the CIR sweeps in front
of the photosphere breaks symmetry through enhanced and asymmetric
absorption of the background limb polarization profile in the wind.
As proof-of-concept, \cite{2023MNRAS.526.1298I} used a conical CIR
to demonstrate the three main observed properties: (a) line effect
for redshifts; (b) the $q-u$ ``blue '' loop for blueshifts; and (c)
the sense of rotation through the loop can be clockwise or
counterclockwise.

\subsection{Colliding Wind Binaries}

Massive colliding wind binaries have drawn considerable interest
from observers and theorists alike.  The effects of the wind collision
lead to significant variations with orbital phase in spectral lines
and continuum across essentially all wavebands, and can be
used to infer wind mass-loss rates, constrain orbital inclination
with ramifications for determining component masses, and probe the
physics of shocks.

Several brighter systems have been studied with polarization, often
interpreted using \cite{1978A&A....68..415B}.  Some studies have
combined colliding wind simulations with radiative transfer
calculations in detailed studies of specific systems
\citep{2002A&A...388..957K}.  To facilitate faster parameter studies,
\cite{2022ApJ...933....5I} adopted the semi-analytic solution of
\cite{1996ApJ...469..729C} for a radiatively cooled bowshock.  The
solution assumes asymptotic terminal speed flow, so the stagnation
point should lie outside the respective wind acceleration zones,
yet remain close enough for radiative cooling to dominate adiabatic
cooling.  The \cite{1996ApJ...469..729C} solution yields the location
of the stagnation point between the stars, the shape of the bowshock,
and its surface density.

Using thin scattering, \cite{2022ApJ...933....5I} showed how this
solution can be used to predict two polarization scale parameters:
$p_1$ for the primary and $p_2$ for the secondary.  The two winds
are assumed spherically symmetric.  Two things break the symmetry
of the problem:  redistribution of material by the wind collision,
and illumination of the medium by two sources.  Then $p_1$ is the
result of scattering polarization by the primary shining on its own
wind that is no longer spherical, on the bowshock structure, and
on the wind of the secondary which of course is not centered on the
primary; then vice versa for $p_2$.  These are purely geometric
parameters predicted by the model.  Cyclic varibility arises from
the viewing inclination, now defined by the tilt of the line-of-centers
between the stars to the observer sightline.

Importantly, colliding wind binaries offer the expectation of
chromatic effects for the polarization because the two stars generally
have distinct spectra.  For binaries involving a WR component with
an OB component, the bowshock will be closer to the OB.  In wavebands
where the WR is the brighter source (EUV and IR), the polarization
will be reduced, but for wavebands where the OB star is brighter,
polarization will be higher \citep[c.f.,][]{2022Ap&SS.367..118S}.
Observing the variable polarization with orbital phase and wavelength
allows to probe the location of the bowshock in the system and
disentangle the relative contributions of the stellar components
to the scattered light.

%
%
%
%

\section{Summary}
\label{sec:summary}

Three different applications for linear polarization as a probe of
wind structure were considered:  clumpy winds, CIRs, and colliding
winds.  For clumpy winds considerations of polarimetric fluctuations
provide additional constraints for interpretive modeling.  A
parameteric clump brightness and polarization profile were utilize
to reproduce the statistical behavior of the observations.  However,
the forms cannot be explained in terms of the simplest models
involving thin scattering and better radiative transfer approaches
for clumps are needed.

The blue loops observed across the He{\sc ii} 4686~\AA\ emission
line in WR~6 can be explained in terms of differential absorption
by a CIR.  A more strategically defined dataset is needed to test
the concept.  Although the simplistic conical CIR model is qualitatively
promising, further modeling of CIRs in WR winds will be needed.

New results for modeling orbital variations of polarization from
colliding winds were described.  These take advantage of the
semi-analytic solution for bowshock shape and surface density from
\cite{1996ApJ...469..729C}.  The results highlight wavelength-dependent
effects arising from the fact that the scattering environment is
illuminated by two different stellar spectral components.  However
the Canto solution applies to radiative cooling, and extending the
model to adiabatic shocks would be valuable.

Massive stars are predominantly UV sources, and although Thomson
scattering is a gray opacity, there are processes that introduce
wavelength-dependent effects that can alter the polarization.
Examples include binarity (as just mentioned), resonance line
scattering, and various opacity effects such as line blanketing for
appropriate cases.  {\em Polstar} is a UV spectropolarimetry concept
that would open a new window into massive star research.  The design
would provide spectroscopy from 115~nm--286~nm, polarimetry from
122~nm--286~nm, a resolving power of $R=20,000$ (velocity resolution
of 15 km/s), and a polarimetric precision of 0.03\%, using a 40~cm
aperture telescope \citep[more details can be found at][Ignace 2024,
in prep]{2022Ap&SS.367..126S}.


%
\acknowledgements{RI gratefully acknowledge support from the National
Science Foundation under grant number AST-2009412.

}

%
\bibliographystyle{stanfest_bibstyle}

{\small 
\bibliography{ignace}

@BOOK{2010stpo.book.....C,
       author = {{Clarke}, David},
        title = "{Stellar Polarimetry}",
         year = 2010,
       adsurl = {https://ui.adsabs.harvard.edu/abs/2010stpo.book.....C}
}

@BOOK{2004ASSL..307.....L,
       author = {{Landi Degl'Innocenti}, E. and {Landolfi}, M.},
        title = "{Polarization in Spectral Lines}",
         year = 2004,
       volume = {307},
          doi = {10.1007/978-1-4020-2415-3},
       adsurl = {https://ui.adsabs.harvard.edu/abs/2004ASSL..307.....L}
}

@BOOK{1960ratr.book.....C,
       author = {{Chandrasekhar}, Subrahmanyan},
        title = "{Radiative transfer}",
         year = 1960,
       adsurl = {https://ui.adsabs.harvard.edu/abs/1960ratr.book.....C}
}

@BOOK{2014tsa..book.....H,
       author = {{Hubeny}, Ivan and {Mihalas}, Dimitri},
        title = "{Theory of Stellar Atmospheres}",
         year = 2014,
       adsurl = {https://ui.adsabs.harvard.edu/abs/2014tsa..book.....H}
}

@ARTICLE{1982MNRAS.200...91S,
       author = {{Simmons}, J.~F.~L.},
        title = "{Analytic treatment of polarization by arbitrary scattering mechanisms in circumstellar envelopes. I - Single stars}",
      journal = {\mnras},
         year = 1982,
        month = jul,
       volume = {200},
        pages = {91-113},
          doi = {10.1093/mnras/200.1.91},
       adsurl = {https://ui.adsabs.harvard.edu/abs/1982MNRAS.200...91S}
}

@ARTICLE{1983MNRAS.205..153S,
       author = {{Simmons}, J.~F.~L.},
        title = "{Analytic treatment of polarization by arbitrary scattering mechanisms in circumstellar envelopes. II - Binary stars}",
      journal = {\mnras},
         year = 1983,
        month = oct,
       volume = {205},
        pages = {153-170},
          doi = {10.1093/mnras/205.1.153},
       adsurl = {https://ui.adsabs.harvard.edu/abs/1983MNRAS.205..153S}
}

@ARTICLE{1999A&A...347..919A,
       author = {{Al-Malki}, M.~B. and {Simmons}, J.~F.~L. and {Ignace}, R. and {Brown}, J.~C. and {Clarke}, D.},
        title = "{Scattering polarization due to light source anisotropy. I. Large spherical envelope}",
      journal = {\aap},
         year = 1999,
        month = jul,
       volume = {347},
        pages = {919-926},
       adsurl = {https://ui.adsabs.harvard.edu/abs/1999A&A...347..919A}
}

@ARTICLE{2009A&A...496..503I,
       author = {{Ignace}, R. and {Al-Malki}, M.~B. and {Simmons}, J.~F.~L. and {Brown}, J.~C. and {Clarke}, D. and {Carson}, J.~C.},
        title = "{Scattering polarization due to light source anisotropy. II. Envelope of arbitrary shape}",
      journal = {\aap},
         year = 2009,
        month = mar,
       volume = {496},
       number = {2},
        pages = {503-511},
          doi = {10.1051/0004-6361:200811214},
       adsurl = {https://ui.adsabs.harvard.edu/abs/2009A&A...496..503I}
}

@ARTICLE{1978A&A....68..415B,
       author = {{Brown}, J.~C. and {McLean}, I.~S. and {Emslie}, A.~G.},
        title = "{Polarisation by Thomson scattering in optically thin stellar envelopes. II. Binary and multiple star envelopes and the determination of binary inclinations.}",
      journal = {\aap},
         year = 1978,
        month = aug,
       volume = {68},
        pages = {415-427},
       adsurl = {https://ui.adsabs.harvard.edu/abs/1978A&A....68..415B}
}

@ARTICLE{1977A&A....57..141B,
       author = {{Brown}, J.~C. and {McLean}, I.~S.},
        title = "{Polarisation by Thomson Scattering in Optically Thin Stellar Envelopes. I. Source Star at Centre of Axisymmetric Envelope}",
      journal = {\aap},
         year = 1977,
        month = may,
       volume = {57},
        pages = {141},
       adsurl = {https://ui.adsabs.harvard.edu/abs/1977A&A....57..141B}
}

@ARTICLE{2013ApJ...766L...9C,
       author = {{Carciofi}, A.~C. and {Faes}, D.~M. and {Townsend}, R.~H.~D. and {Bjorkman}, J.~E.},
        title = "{Polarimetric Observations of {\ensuremath{\sigma}} Orionis E}",
      journal = {\apjl},
         year = 2013,
        month = mar,
       volume = {766},
       number = {1},
          eid = {L9},
        pages = {L9},
          doi = {10.1088/2041-8205/766/1/L9},
archivePrefix = {arXiv},
       eprint = {1302.4684},
 primaryClass = {astro-ph.SR},
       adsurl = {https://ui.adsabs.harvard.edu/abs/2013ApJ...766L...9C}
}

@ARTICLE{1996A&A...306..519R,
       author = {{Richardson}, L.~L. and {Brown}, J.~C. and {Simmons}, J.~F.~L.},
        title = "{Polarimetric versus photometric variability and the density of WR star wind inhomogeneities.}",
      journal = {\aap},
         year = 1996,
        month = feb,
       volume = {306},
        pages = {519},
       adsurl = {https://ui.adsabs.harvard.edu/abs/1996A&A...306..519R}
}

@ARTICLE{2000A&A...357..233L,
       author = {{Li}, Q. and {Brown}, J.~C. and {Ignace}, R. and {Cassinelli}, J.~P. and {Oskinova}, L.~M.},
        title = "{Wolf-Rayet wind structure and optical variability}",
      journal = {\aap},
         year = 2000,
        month = may,
       volume = {357},
        pages = {233-240},
       adsurl = {https://ui.adsabs.harvard.edu/abs/2000A&A...357..233L}
}

@ARTICLE{2015A&A...575A.129I,
       author = {{Ignace}, Richard and {St-Louis}, Nicole and {Proulx-Giraldeau}, F{\'e}lix},
        title = "{Polarimetric modeling of corotating interaction regions threading massive-star winds}",
      journal = {\aap},
         year = 2015,
        month = mar,
       volume = {575},
          eid = {A129},
        pages = {A129},
          doi = {10.1051/0004-6361/201424806},
archivePrefix = {arXiv},
       eprint = {1501.07563},
 primaryClass = {astro-ph.SR},
       adsurl = {https://ui.adsabs.harvard.edu/abs/2015A&A...575A.129I}
}

@ARTICLE{1987ApJ...317..290C,
       author = {{Cassinelli}, J.~P. and {Nordsieck}, K.~H. and {Murison}, M.~A.},
        title = "{Polarization of Light Scattered from the Winds of Early-Type Stars}",
      journal = {\apj},
         year = 1987,
        month = jun,
       volume = {317},
        pages = {290},
          doi = {10.1086/165277},
       adsurl = {https://ui.adsabs.harvard.edu/abs/1987ApJ...317..290C}
}

@ARTICLE{1989ApJ...344..341B,
       author = {{Brown}, John C. and {Carlaw}, Vivette A. and {Cassinelli}, Joseph P.},
        title = "{Finite Source Depolarization Factors for Circumstellar Scattering}",
      journal = {\apj},
         year = 1989,
        month = sep,
       volume = {344},
        pages = {341},
          doi = {10.1086/167803},
       adsurl = {https://ui.adsabs.harvard.edu/abs/1989ApJ...344..341B}
}

@ARTICLE{1994ApJ...436..818B,
       author = {{Bjorkman}, J.~E. and {Bjorkman}, K.~S.},
        title = "{The Effects of Gravity Darkening on the Ultraviolet Continuum Polarization Produced by Circumstellar Disks}",
      journal = {\apj},
         year = 1994,
        month = dec,
       volume = {436},
        pages = {818},
          doi = {10.1086/174958},
       adsurl = {https://ui.adsabs.harvard.edu/abs/1994ApJ...436..818B}
}

@ARTICLE{2007ARA&A..45..177C,
       author = {{Crowther}, Paul A.},
        title = "{Physical Properties of Wolf-Rayet Stars}",
      journal = {\araa},
         year = 2007,
        month = sep,
       volume = {45},
       number = {1},
        pages = {177-219},
          doi = {10.1146/annurev.astro.45.051806.110615},
archivePrefix = {arXiv},
       eprint = {astro-ph/0610356},
 primaryClass = {astro-ph},
       adsurl = {https://ui.adsabs.harvard.edu/abs/2007ARA&A..45..177C}
}

@ARTICLE{2012ARA&A..50..107L,
       author = {{Langer}, N.},
        title = "{Presupernova Evolution of Massive Single and Binary Stars}",
      journal = {\araa},
         year = 2012,
        month = sep,
       volume = {50},
        pages = {107-164},
          doi = {10.1146/annurev-astro-081811-125534},
archivePrefix = {arXiv},
       eprint = {1206.5443},
 primaryClass = {astro-ph.SR},
       adsurl = {https://ui.adsabs.harvard.edu/abs/2012ARA&A..50..107L}
}

@ARTICLE{2008A&ARv..16..209P,
       author = {{Puls}, Joachim and {Vink}, Jorick S. and {Najarro}, Francisco},
        title = "{Mass loss from hot massive stars}",
      journal = {\aapr},
         year = 2008,
        month = dec,
       volume = {16},
       number = {3-4},
        pages = {209-325},
          doi = {10.1007/s00159-008-0015-8},
archivePrefix = {arXiv},
       eprint = {0811.0487},
 primaryClass = {astro-ph},
       adsurl = {https://ui.adsabs.harvard.edu/abs/2008A&ARv..16..209P}
}

@ARTICLE{1987ApJ...322..870S,
       author = {{St.-Louis}, Nicole and {Drissen}, Laurent and {Moffat}, Anthony F.~J. and {Bastien}, Pierre and {Tapia}, Santiago},
        title = "{Polarization Variability among Wolf-Rayet Stars. I. Linear Polarization of a Complete Sample of Southern Galactic WC Stars}",
      journal = {\apj},
         year = 1987,
        month = nov,
       volume = {322},
        pages = {870},
          doi = {10.1086/165782},
       adsurl = {https://ui.adsabs.harvard.edu/abs/1987ApJ...322..870S}
}

@ARTICLE{1987ApJ...322..888D,
       author = {{Drissen}, Laurent and {St. -Louis}, Nicole and {Moffat}, Anthony F.~J. and {Bastien}, Pierre},
        title = "{Polarization Variability among Wolf-Rayet Stars. II. Linear Polarization as a Complete Sample of Southern Galactic WN Stars}",
      journal = {\apj},
         year = 1987,
        month = nov,
       volume = {322},
        pages = {888},
          doi = {10.1086/165783},
       adsurl = {https://ui.adsabs.harvard.edu/abs/1987ApJ...322..888D}
}

@ARTICLE{1995A&A...295..725B,
       author = {{Brown}, J.~C. and {Richardson}, L.~L. and {Antokhin}, I. and {Robert}, C. and {Moffat}, A.~F.~J. and {St-Louis}, N.},
        title = "{Combined spectrometric, photometric and polarimetric diagnostics for `blobs' in WR star winds.}",
      journal = {\aap},
         year = 1995,
        month = mar,
       volume = {295},
        pages = {725},
       adsurl = {https://ui.adsabs.harvard.edu/abs/1995A&A...295..725B}
}

@ARTICLE{1995ApJ...441..400C,
       author = {{Code}, Arthur D. and {Whitney}, Barbara A.},
        title = "{Polarization from Scattering in Blobs}",
      journal = {\apj},
         year = 1995,
        month = mar,
       volume = {441},
        pages = {400},
          doi = {10.1086/175363},
       adsurl = {https://ui.adsabs.harvard.edu/abs/1995ApJ...441..400C}
}

@ARTICLE{2000ApJ...540..412R,
       author = {{Rodrigues}, Cl{\'a}udia V. and {Magalh{\~a}es}, A. M{\'a}rio},
        title = "{Blobs in Wolf-Rayet Winds: Random Photometric and Polarimetric Variability}",
      journal = {\apj},
         year = 2000,
        month = sep,
       volume = {540},
       number = {1},
        pages = {412-421},
          doi = {10.1086/309291},
archivePrefix = {arXiv},
       eprint = {astro-ph/0003362},
 primaryClass = {astro-ph},
       adsurl = {https://ui.adsabs.harvard.edu/abs/2000ApJ...540..412R}
}

@INPROCEEDINGS{2012AIPC.1429..278T,
       author = {{Townsend}, Rich},
        title = "{A Monte-Carlo method for simulating linear polarization variations in clumpy massive-star winds}",
    booktitle = {Stellar Polarimetry: from Birth to Death},
         year = 2012,
       editor = {{Hoffman}, Jennifer L. and {Bjorkman}, Jon and {Whitney}, Barbara},
       series = {American Institute of Physics Conference Series},
       volume = {1429},
        month = may,
    publisher = {AIP},
        pages = {278-281},
          doi = {10.1063/1.3701941},
       adsurl = {https://ui.adsabs.harvard.edu/abs/2012AIPC.1429..278T}
}

@ARTICLE{2007A&A...469.1045D,
       author = {{Davies}, B. and {Vink}, J.~S. and {Oudmaijer}, R.~D.},
        title = "{Modelling the clumping-induced polarimetric variability of hot star winds}",
      journal = {\aap},
         year = 2007,
        month = jul,
       volume = {469},
       number = {3},
        pages = {1045-1056},
          doi = {10.1051/0004-6361:20077193},
archivePrefix = {arXiv},
       eprint = {0704.2569},
 primaryClass = {astro-ph},
       adsurl = {https://ui.adsabs.harvard.edu/abs/2007A&A...469.1045D}
}

@ARTICLE{2019MNRAS.490.5921R,
       author = {{Ramiaramanantsoa}, Tahina and {Ignace}, Richard and {Moffat}, Anthony F.~J. and {St-Louis}, Nicole and {Shkolnik}, Evgenya L. and {Popowicz}, Adam and {Kuschnig}, Rainer and {Pigulski}, Andrzej and {Wade}, Gregg A. and {Handler}, Gerald and {Pablo}, Herbert and {Zwintz}, Konstanze},
        title = "{The chaotic wind of WR 40 as probed by BRITE}",
      journal = {\mnras},
         year = 2019,
        month = dec,
       volume = {490},
       number = {4},
        pages = {5921-5930},
          doi = {10.1093/mnras/stz2895},
archivePrefix = {arXiv},
       eprint = {1910.05258},
 primaryClass = {astro-ph.SR},
       adsurl = {https://ui.adsabs.harvard.edu/abs/2019MNRAS.490.5921R}
}

@ARTICLE{2023MNRAS.519.3271I,
       author = {{Ignace}, R. and {Moffat}, A.~F.~J. and {Robert}, C. and {Drissen}, L.},
        title = "{Constraints on clumps in the representative wind of the WN8 Wolf-Rayet star HD 96548 = WR 40 with simultaneous broad-band light and linear-polarization variability}",
      journal = {\mnras},
         year = 2023,
        month = mar,
       volume = {519},
       number = {3},
        pages = {3271-3280},
          doi = {10.1093/mnras/stac3772},
archivePrefix = {arXiv},
       eprint = {2212.13858},
 primaryClass = {astro-ph.SR},
       adsurl = {https://ui.adsabs.harvard.edu/abs/2023MNRAS.519.3271I}
}

@ARTICLE{1976ApJS...32..715L,
       author = {{Lamers}, H.~J.~G.~L.~M. and {Morton}, D.~C.},
        title = "{Mass ejection from the O4f star Zeta Puppis.}",
      journal = {\apjs},
         year = 1976,
        month = oct,
       volume = {32},
        pages = {715-736},
          doi = {10.1086/190413},
       adsurl = {https://ui.adsabs.harvard.edu/abs/1976ApJS...32..715L}
}

@ARTICLE{1999A&A...344..231K,
       author = {{Kaper}, L. and {Henrichs}, H.~F. and {Nichols}, J.~S. and {Telting}, J.~H.},
        title = "{Long- and short-term variability in O-star winds. II. Quantitative analysis of DAC behaviour}",
      journal = {\aap},
         year = 1999,
        month = apr,
       volume = {344},
        pages = {231-262},
          doi = {10.48550/arXiv.astro-ph/9812427},
archivePrefix = {arXiv},
       eprint = {astro-ph/9812427},
 primaryClass = {astro-ph},
       adsurl = {https://ui.adsabs.harvard.edu/abs/1999A&A...344..231K}
}

@ARTICLE{2015ApJ...809...12M,
       author = {{Massa}, D. and {Prinja}, R.~K.},
        title = "{On the Origin of Wind Line Variability in O Stars}",
      journal = {\apj},
         year = 2015,
        month = aug,
       volume = {809},
       number = {1},
          eid = {12},
        pages = {12},
          doi = {10.1088/0004-637X/809/1/12},
archivePrefix = {arXiv},
       eprint = {1506.06605},
 primaryClass = {astro-ph.SR},
       adsurl = {https://ui.adsabs.harvard.edu/abs/2015ApJ...809...12M}
}

@ARTICLE{1984ApJ...283..303M,
       author = {{Mullan}, D.~J.},
        title = "{Corotating interaction regions in stellar winds}",
      journal = {\apj},
         year = 1984,
        month = aug,
       volume = {283},
        pages = {303-312},
          doi = {10.1086/162307},
       adsurl = {https://ui.adsabs.harvard.edu/abs/1984ApJ...283..303M}
}

@ARTICLE{1996ApJ...462..469C,
       author = {{Cranmer}, Steven R. and {Owocki}, Stanley P.},
        title = "{Hydrodynamical Simulations of Corotating Interaction Regions and Discrete Absorption Components in Rotating O-Star Winds}",
      journal = {\apj},
         year = 1996,
        month = may,
       volume = {462},
        pages = {469},
          doi = {10.1086/177166},
archivePrefix = {arXiv},
       eprint = {astro-ph/9508004},
 primaryClass = {astro-ph},
       adsurl = {https://ui.adsabs.harvard.edu/abs/1996ApJ...462..469C}
}

@ARTICLE{2008ApJ...678..408L,
       author = {{Lobel}, A. and {Blomme}, R.},
        title = "{Modeling Ultraviolet Wind Line Variability in Massive Hot Stars}",
      journal = {\apj},
         year = 2008,
        month = may,
       volume = {678},
       number = {1},
        pages = {408-430},
          doi = {10.1086/529129},
archivePrefix = {arXiv},
       eprint = {0712.3804},
 primaryClass = {astro-ph},
       adsurl = {https://ui.adsabs.harvard.edu/abs/2008ApJ...678..408L}
}

@ARTICLE{2004A&A...423..693D,
       author = {{Dessart}, L.},
        title = "{3D hydrodynamical simulations of corotating interaction regions in rotating line-driven stellar winds}",
      journal = {\aap},
         year = 2004,
        month = aug,
       volume = {423},
        pages = {693-704},
          doi = {10.1051/0004-6361:20040543},
       adsurl = {https://ui.adsabs.harvard.edu/abs/2004A&A...423..693D}
}

@ARTICLE{2017MNRAS.470.3672D,
       author = {{David-Uraz}, A. and {Owocki}, S.~P. and {Wade}, G.~A. and {Sundqvist}, J.~O. and {Kee}, N.~D.},
        title = "{Investigating the origin of cyclical wind variability in hot massive stars - II. Hydrodynamical simulations of corotating interaction regions using realistic spot parameters for the O giant {\ensuremath{\xi}} Persei}",
      journal = {\mnras},
         year = 2017,
        month = sep,
       volume = {470},
       number = {3},
        pages = {3672-3684},
          doi = {10.1093/mnras/stx1478},
archivePrefix = {arXiv},
       eprint = {1706.03647},
 primaryClass = {astro-ph.SR},
       adsurl = {https://ui.adsabs.harvard.edu/abs/2017MNRAS.470.3672D}
}

@ARTICLE{2009ApJ...698.1951S,
       author = {{St-Louis}, N. and {Chen{\'e}}, A. -N. and {Schnurr}, O. and {Nicol}, M. -H.},
        title = "{A Systematic Search for Corotating Interaction Regions in Apparently Single Galactic Wolf-Rayet Stars. I. Characterizing the Variability}",
      journal = {\apj},
         year = 2009,
        month = jun,
       volume = {698},
       number = {2},
        pages = {1951-1962},
          doi = {10.1088/0004-637X/698/2/1951},
archivePrefix = {arXiv},
       eprint = {0905.4460},
 primaryClass = {astro-ph.SR},
       adsurl = {https://ui.adsabs.harvard.edu/abs/2009ApJ...698.1951S}
}

@ARTICLE{2011ApJ...735...34C,
       author = {{Chen{\'e}}, A. -N. and {Moffat}, A.~F.~J. and {Cameron}, C. and {Fahed}, R. and {Gamen}, R.~C. and {Lef{\`e}vre}, L. and {Rowe}, J.~F. and {St-louis}, N. and {Muntean}, V. and {De La Chevroti{\`e}re}, A. and {Guenther}, D.~B. and {Kuschnig}, R. and {Matthews}, J.~M. and {Rucinski}, S.~M. and {Sasselov}, D. and {Weiss}, W.~W.},
        title = "{WR 110: A Single Wolf-Rayet Star with Corotating Interaction Regions in its Wind?}",
      journal = {\apj},
         year = 2011,
        month = jul,
       volume = {735},
       number = {1},
          eid = {34},
        pages = {34},
          doi = {10.1088/0004-637X/735/1/34},
archivePrefix = {arXiv},
       eprint = {1105.0919},
 primaryClass = {astro-ph.SR},
       adsurl = {https://ui.adsabs.harvard.edu/abs/2011ApJ...735...34C}
}

@ARTICLE{2011ApJ...736..140C,
       author = {{Chen{\'e}}, A. -N. and {St-Louis}, N.},
        title = "{A Systematic Search for Corotating Interaction Regions in Apparently Single Galactic Wolf-Rayet Stars. II. A Global View of the Wind Variability}",
      journal = {\apj},
         year = 2011,
        month = aug,
       volume = {736},
       number = {2},
          eid = {140},
        pages = {140},
          doi = {10.1088/0004-637X/736/2/140},
archivePrefix = {arXiv},
       eprint = {1105.5133},
 primaryClass = {astro-ph.SR},
       adsurl = {https://ui.adsabs.harvard.edu/abs/2011ApJ...736..140C}
}

@ARTICLE{2016MNRAS.460.3407A,
       author = {{Aldoretta}, E.~J. and {St-Louis}, N. and {Richardson}, N.~D. and {Moffat}, A.~F.~J. and {Eversberg}, T. and {Hill}, G.~M. and {Shenar}, T. and {Artigau}, {\'E}. and {Gauza}, B. and {Knapen}, J.~H. and {Kub{\'a}t}, J. and {Kub{\'a}tov{\'a}}, B. and {Maltais-Tariant}, R. and {Mu{\~n}oz}, M. and {Pablo}, H. and {Ramiaramanantsoa}, T. and {Richard-Laferri{\`e}re}, A. and {Sablowski}, D.~P. and {Sim{\'o}n-D{\'\i}az}, S. and {St-Jean}, L. and {Bolduan}, F. and {Dias}, F.~M. and {Dubreuil}, P. and {Fuchs}, D. and {Garrel}, T. and {Grutzeck}, G. and {Hunger}, T. and {K{\"u}sters}, D. and {Langenbrink}, M. and {Leadbeater}, R. and {Li}, D. and {Lopez}, A. and {Mauclaire}, B. and {Moldenhawer}, T. and {Potter}, M. and {dos Santos}, E.~M. and {Schanne}, L. and {Schmidt}, J. and {Sieske}, H. and {Strachan}, J. and {Stinner}, E. and {Stinner}, P. and {Stober}, B. and {Strandbaek}, K. and {Syder}, T. and {Verilhac}, D. and {Waldschl{\"a}ger}, U. and {Weiss}, D. and {Wendt}, A.},
        title = "{An extensive spectroscopic time series of three Wolf-Rayet stars - I. The lifetime of large-scale structures in the wind of WR 134}",
      journal = {\mnras},
         year = 2016,
        month = aug,
       volume = {460},
       number = {3},
        pages = {3407-3417},
          doi = {10.1093/mnras/stw1188},
archivePrefix = {arXiv},
       eprint = {1605.04868},
 primaryClass = {astro-ph.SR},
       adsurl = {https://ui.adsabs.harvard.edu/abs/2016MNRAS.460.3407A}
}

@ARTICLE{2023MNRAS.526.1298I,
       author = {{Ignace}, R. and {Bjorkman}, J.~E. and {Chen{\'e}}, A. -N. and {Erba}, C. and {Fabiani}, L. and {Moffat}, A.~F.~J. and {Sincennes}, R. and {St-Louis}, N.},
        title = "{Modelling variable linear polarization produced by Co-rotating Interaction Regions (CIRs) across optical recombination lines of Wolf-Rayet stars}",
      journal = {\mnras},
         year = 2023,
        month = nov,
       volume = {526},
       number = {1},
        pages = {1298-1307},
          doi = {10.1093/mnras/stad2878},
archivePrefix = {arXiv},
       eprint = {2309.10599},
 primaryClass = {astro-ph.SR},
       adsurl = {https://ui.adsabs.harvard.edu/abs/2023MNRAS.526.1298I}
}

@ARTICLE{2002A&A...388..957K,
       author = {{Kurosawa}, R. and {Hillier}, D.~J. and {Pittard}, J.~M.},
        title = "{Mass-loss rate determination for the massive binary V444 Cygni using 3-D Monte-Carlo simulations of line and polarization variability}",
      journal = {\aap},
         year = 2002,
        month = jun,
       volume = {388},
        pages = {957-977},
          doi = {10.1051/0004-6361:20020443},
archivePrefix = {arXiv},
       eprint = {astro-ph/0203451},
 primaryClass = {astro-ph},
       adsurl = {https://ui.adsabs.harvard.edu/abs/2002A&A...388..957K}
}

@ARTICLE{1996ApJ...469..729C,
       author = {{Cant{\'o}}, J. and {Raga}, A.~C. and {Wilkin}, F.~P.},
        title = "{Exact, Algebraic Solutions of the Thin-Shell Two-Wind Interaction Problem}",
      journal = {\apj},
         year = 1996,
        month = oct,
       volume = {469},
        pages = {729},
          doi = {10.1086/177820},
       adsurl = {https://ui.adsabs.harvard.edu/abs/1996ApJ...469..729C}
}
}

\end{document}